\documentclass[twocolumn]{aastex62}
\usepackage{tabularx}
\newcolumntype{Y}{>{\centering\arraybackslash}X}

\graphicspath{{./}{figures/}}

\begin{document}
\title{Beyond Gaia: Asteroseismic Distances of M giants using Ground-Based Transient Surveys}

\author[0000-0002-5504-8752]{Connor Auge}
\affiliation{Institute for Astronomy, University of Hawai\`{}i, 2680 Woodlawn Drive, Honolulu, HI 96822, USA}

\correspondingauthor{Connor Auge}
\email{cauge@hawaii.edu}

\author[0000-0001-8832-4488]{Daniel Huber}
\affiliation{Institute for Astronomy, University of Hawai\`{}i, 2680 Woodlawn Drive, Honolulu, HI 96822, USA}

\author{Aren Heinze}
\affiliation{Institute for Astronomy, University of Hawai\`{}i, 2680 Woodlawn Drive, Honolulu, HI 96822, USA}

\author[0000-0003-4631-1149]{B.~J.~Shappee}
\affiliation{Institute for Astronomy, University of Hawai\`{}i, 2680 Woodlawn Drive, Honolulu, HI 96822, USA}

\author{John Tonry}
\affiliation{Institute for Astronomy, University of Hawai\`{}i, 2680 Woodlawn Drive, Honolulu, HI 96822, USA}

\author{Sukanya Chakrabarti}
\affiliation{School  of  Physics  and  Astronomy,  Rochester  Institute  of Technology,  84  Lomb  Memorial  Drive,  Rochester,  NY  14623}

\author{Robyn E. Sanderson}
\affiliation{Department of Physics \& Astronomy, University of Pennsylvania, 209 South 33rd Street, Philadelphia, PA 19104 USA}
\affiliation{Center for Computational Astrophysics, Flatiron Institute, 162 Fifth Avenue, New York, NY 10010 USA} 

\author{Larry Denneau}
\affiliation{Institute for Astronomy, University of Hawai\`{}i, 2680 Woodlawn Drive, Honolulu, HI 96822, USA}

\author{Heather Flewelling}
\affiliation{Institute for Astronomy, University of Hawai\`{}i, 2680 Woodlawn Drive, Honolulu, HI 96822, USA}

\author[0000-0001-9206-3460]{Thomas~W.-S.~Holoien}
\altaffiliation{Carnegie Fellow}
\affiliation{The Observatories of the Carnegie Institution for Science, 813 Santa Barbara St., Pasadena, CA 91101, USA}

\author{C. S. Kochanek}
\affiliation{Department of Astronomy, The Ohio State University, 140 West 18th Avenue, Columbus, OH 43210, USA}
\affiliation{Center for Cosmology and Astroparticle Physics, The Ohio State University, 191 W.~Woodruff Avenue, Columbus, OH 43210, USA}

\author{Giuliano Pignata}
\affiliation{Departamento de Ciencias Fisicas, Universidad Andres Bello, Avda. Republica 252, Santiago, Chile}
\affiliation{Millennium Institute of Astrophysics (MAS), Nuncio Monsenor Śotero Sanz 100, Providencia,Santiago, Chile}

\author{Amanda Sickafoose}
\affiliation{Planetary Sciences Institute, 1700 East Fort Lowell, Tucson, AZ 85719}
\affiliation{Department of Earth, Atmospheric, and Planetary Sciences, Massachusetts Institute of Technology, 77 Massachusetts Ave., Cambridge, MA 02139}

\author{Brian Stalder}
\affiliation{Rubin Observatory Project Office, 950 N Cherry Ave, Tucson, AZ, USA 85719}

\author{K.Z. Stanek}
\affiliation{Department of Astronomy, The Ohio State University, 140 West 18th Avenue, Columbus, OH 43210, USA}
\affiliation{Center for Cosmology and Astroparticle Physics, The Ohio State University, 191 W.~Woodruff Avenue, Columbus, OH 43210, USA}

\author{Dennis Stello}
\affiliation{School of Physics, University of New South Wales, NSW 2052, Australia}
\affiliation{Sydney Institute for Astronomy (SIfA), school of Physics, University of Sydney, NSW 2006 Australia}
\affiliation{Department of Physics and Astronomy, Stellar Astrophysics Centre, Aarhus University, DK-8000 Aarhus, C, Denmark}

\author[0000-0003-2377-9574]{Todd A. Thompson}
\affiliation{Department of Astronomy, The Ohio State University, 140 West 18th Avenue, Columbus, OH 43210, USA}
\affiliation{Center for Cosmology and Astroparticle Physics, The Ohio State University, 191 W.~Woodruff Avenue, Columbus, OH 43210, USA}

\begin{abstract}
    Evolved stars near the tip of the red giant branch (TRGB) show solar-like oscillations with periods spanning hours to months and amplitudes ranging from $\sim$1 $\rm{mmag}$ to $\sim$100 $\rm{mmag}$. The systematic detection of the resulting photometric variations with ground-based telescopes would enable the application of asteroseismology to a much larger and more distant sample of stars than is currently accessible with space-based telescopes such as \textit{Kepler} or the ongoing Transiting Exoplanet Survey Satellite (\textit{TESS}) mission. We present an asteroseismic analysis of 493 M giants using data from two ground-based surveys: the Asteroid Terrestrial-impact Last Alert System (ATLAS) and the All-Sky Automated Survey for Supernovae (ASAS-SN). By comparing the extracted frequencies with constraints from \textit{Kepler}, the Sloan Digital Sky Survey Apache Point Observatory Galaxy Evolution Experiment (APOGEE), and Gaia we demonstrate that ground-based transient surveys allow accurate distance measurements to oscillating M giants with a precision of $\sim$15$\%$. Using stellar population synthesis models we predict that ATLAS and ASAS-SN can provide asteroseismic distances to $\sim$2$\times$10$^{6}$ galactic M giants out to typical distances of $20-50 \; \rm{kpc}$, vastly improving the reach of Gaia and providing critical constraints for Galactic archaeology and galactic dynamics.
\end{abstract}


\section{Introduction}

Asteroseismology, the study of stellar structure through the observations of pulsations, was revolutionized by the launch of the \textit{Kepler} space telescope \citep{Borucki2010}. The long-baseline, high-quality photometry enabled an in-depth analysis of stellar oscillations for an unprecedented number of stars. This new wealth of data provided a means to systematically determine fundamental stellar properties for stars at a variety of different evolutionary states \citep[e.g.,][]{Chaplin2013,Garcia2018}. 

One area of study that saw particular success was the analysis of solar-like oscillations in stars evolving up the red giant branch (RGB). Such studies determined the evolutionary stages of K giants \citep[e.g.,][]{Beck2011,Bedding2011,Mosser2012b,Stello_2013}, constrained their internal rotation \citep[e.g.,][]{Beck2012,Deheuvels2012,Mosser2012b,Deheuvels2014}, and characterized  exoplanet properties \citep[e.g.,][]{Huber2013,Quinn2015,Grunblatt2019}. \textit{Kepler}, $K2$ \citep{Howell2014}, and the CoRoT \citep{Baglin2006} space telescope also significantly advanced the field of Galactic archaeology, the study of the structure and evolution of the Milky Way, by examining stellar populations in different parts of the Galaxy \citep[e.g.,][]{Miglio2013,Stello2015,Casagrande2016,Sharma2016,Rendle2019}. These studies were possible thanks to the rich oscillation power spectra of these RGB stars. 

Power spectra for solar-like oscillators are characterized by a Gaussian envelope of oscillation modes. The peak of this envelope is defined as the frequency of maximum power ($\nu_{\mathrm{max}}$) and the average separation between modes of the same spherical degree and consecutive radial order is known as the large frequency separation ($\Delta \nu$). As a star evolves up the RGB, these asteroseismic quantities smoothly shift to lower frequencies and smaller separations due to the expansion in the stellar radius \citep{Hekker2017, Garcia2018}. Photometric studies of M giants \citep[e.g.,][]{Bedding1998,Kiss2003,Kiss2004,Ita2004,Groenewegen2004,Soszynski2007} have shown that this semi-regular variability has the same physical origin as the solar-like oscillations seen in less luminous stars, just shifted in frequency \citep{Tabur2010,Stello2014}. Extensive studies of M giants have provided an in-depth analysis of the different oscillation modes and shown their potential for use as distance indicators \citep[e.g.,][]{Banyai2013,Mosser2013}.

Because asteroseismology precisely constrains fundamental properties such as mass, age, and radius, it can be used as a powerful distance indicator. \cite{Mathur2016} demonstrated that the distance to the high luminosity red giants observed by \textit{Kepler} can be measured with a precision to a few percent out to tens of kiloparsecs. \cite{Huber2017_testingScaling} found that asteroseismology is expected to provide more precise distances than Gaia at the end of its mission for stars beyond $\sim3 \; \rm{kpc}$. While mass and age are more challenging to derive for M giants near the tip of the RGB (TRGB) through asteroseismology, precise distances can be determined for stars outside the reach of Gaia through the use of known period-luminosity relations \citep[discussed further in \S4.1]{Tabur2010}. Stars traditionally used in period-luminosity relations, such as Cepheids and RR Lyrae stars, have been successfully used to map spatially distinct stellar density features out to $\sim60 \; \rm{kpc} - 120 \; \rm{kpc}$ with a precision of $\sim7\%$ \citep{Drake2013,Sesar2017}. However, an even more thorough analysis could be done with distant M giants as they are much more numerous and more luminous than the classical stellar oscillators \citep{Skowron2019}. M giants can also be used in tandem with classical variable stars for a unique analysis of the Milky Way halo. For example, \cite{Price-Whelan2015} and \cite{Sanderson2017} used the relative numbers of RR Lyrae stars and M giants in the outer regions of the Galaxy, to constrain the accretion history of the Milky Way. 

Photometric data from space-based missions such as \textit{Kepler}, $K2$, and the ongoing Transiting Exoplanet Survey Satellite (\textit{TESS}) mission \citep{Ricker2015} are well suited for determining asteroseismic masses and ages of K giants and relatively nearby stellar populations. However, they are not as well suited for studies of the outer regions of the Galaxy. The main \textit{Kepler} mission examined a relatively small region of the sky ($\sim$116 square degrees), which restricts studies to a single line of sight through the Galaxy. The extended $K2$ mission had multiple fields-of-view across the ecliptic plane over the 19 campaigns, with each campaign lasting approximately 80 days \citep{Howell2014}. While $K2$ provided more comprehensive coverage across the Galaxy, the shorter observational baseline poses a challenge to conducting asteroseismology for the most evolved red giants near the TRGB, which have typical oscillation periods greater than 30 days. While \textit{TESS} will provide nearly complete coverage across the entire sky, it will have observational baselines of approximately 30 days for the majority of the observing area. Only near the ecliptic poles in the continuous viewing zones does \textit{TESS} have a long enough baseline to detect the long oscillation periods of evolved red giants. 

\begin{figure}[t!]
    \centering
    \includegraphics[width=\linewidth]{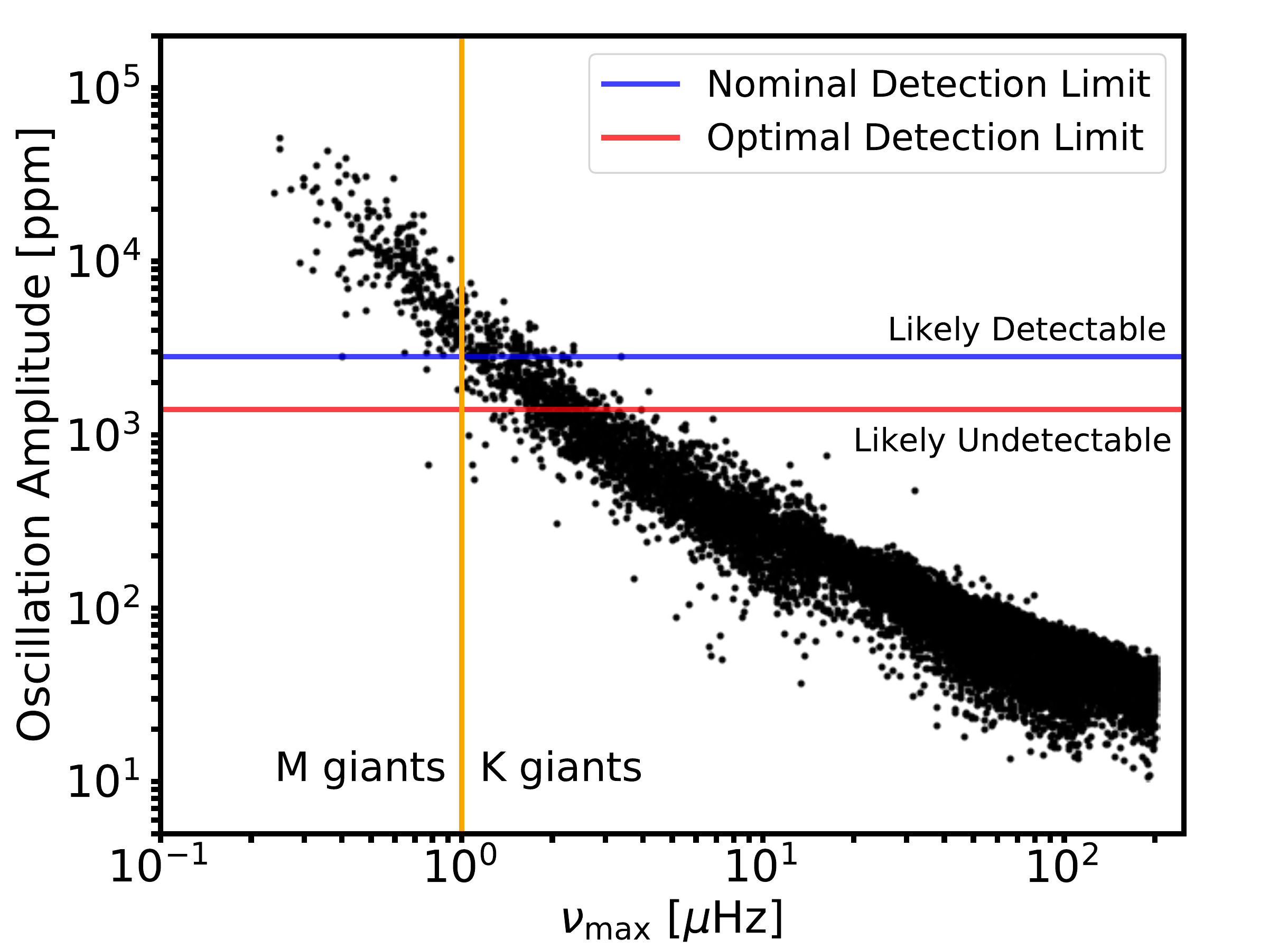}
    \caption{Oscillation amplitude as a function of the frequency of maximum power for \textit{Kepler} stars \citep{Yu2020}. The nominal and optimal photometric precision of the combined ATLAS and ASAS-SN light curves containing 100 data points for a V = 13 magnitude star is shown by the blue and red lines respectively. The vertical line indicates the $\nu_{\mathrm{max}}$ value separating M and K giants.}
    \label{fig:precision}
\end{figure}

Fortunately, oscillation amplitudes increase with the luminosity of the star, and stars near the TRGB show amplitudes on the order of several parts per thousand. This allows the oscillation modes of luminous M giants throughout the Galaxy to be observed by ground-based telescopes, provided the data covers a long enough observational baseline to accurately constrain the period. The growing number of large-scale, ground-based surveys, with years of photometric data covering nearly the entire sky, provide this new means to study stellar variability. Surveys such as the Asteroid Terrestrial-impact Last Alert System \citep[ATLAS,][]{ATLAS_original,Tonry2018}, the All-Sky Automated Survey for Supernovae \citep[ASAS-SN,][]{ASASSN_original,ASASSN_original2}, the Panoramic Survey Telescope and Rapid Response System \citep[Pan-STARRS,][]{Chambers2016}, and the Zwicky Transient Facility \citep[ZTF,][]{Bellm2019} have the photometric precision, observational baseline, and sky coverage to make large scale Galactic archaeology studies possible through the analysis of variable stars. Significant work has already been done to classify variable stars utilizing data from these surveys \citep[e.g.,][]{ATLASvariablestars,ASASSN_variable,Jayasinghe2019_V,Jayasinghe2019_VI,Jayasinghe2019_II,Jayasinghe2019_III,Pawlak2019_IV}, but no in-depth asteroseismic analysis of M giants has been previously conducted. 

Here we provide a proof of concept that these ground-based surveys can be used to systematically perform asteroseismology of M giants and thus lay the foundation for precise distance measurements throughout the galaxy. We do this by using photometry from ATLAS and ASAS-SN to perform asteroseismology on a sample of M giants in the \textit{Kepler} field and then compare the asteroseismic observables determined from ground and space-based observations. We also compare the asteroseismic surface gravities determined with ATLAS and ASAS-SN data to those determined spectroscopically by the Sloan Digital Sky Survey Apache Point Observatory Galaxy Evolution Experiment \citep[APOGEE,][]{Majewski2017} for a sample of stars outside of the \textit{Kepler} field.

\section{Data and Methods}
\subsection{Target Selection}

\begin{figure}[t!]
    \centering
    \includegraphics[width=\linewidth]{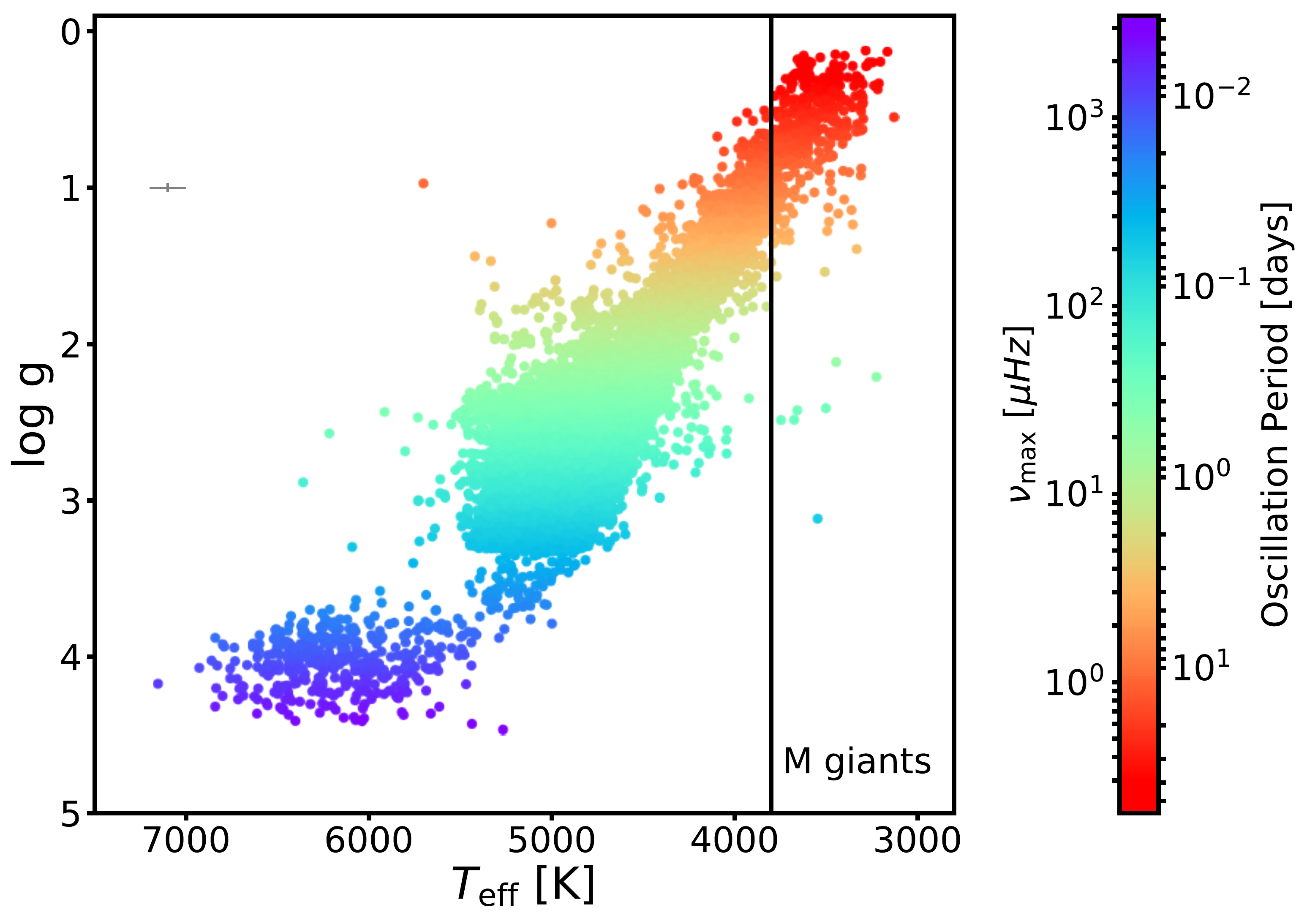}
    \caption{An H-R diagram using surface gravities measured with asteroseismology from \cite{Chaplin2013,Yu2018} and \cite{Yu2020}. A typical error bar is shown in the upper left. Points are colored by the measured frequency of maximum power and the corresponding oscillation period. Stars to the right of the vertical line are M giants. Only the stars in red near the TRGB have large enough oscillation amplitudes to be detected with the currently available ATLAS and ASAS-SN data.}
    \label{fig:HR}
\end{figure}

Amplitudes of solar-like oscillations have been predicted to scale linearly with stellar luminosity \citep{Christensen-Dalsgaard1983, KjeldsenBedding}, and hence should scale inversely with $\nu_{\rm{max}}$. Figure \ref{fig:precision} shows oscillation amplitudes as a function of $\nu_{\rm{max}}$ for a sample of \textit{Kepler} stars measured by \cite{Yu2020}. The nominal, current photometric precision of ATLAS and ASAS-SN light curves containing 100 data points is shown for a star with a V magnitude of $\sim$13. This limit was calculated assuming a photometric error floor of 0.02 magnitudes \citep{Jayasinghe2019_III} and estimating the one sigma uncertainty in the amplitude as
\begin{equation}
    \sigma(a) = \sqrt{\frac{2}{N}}\sigma(m)\:,
    \label{eq:error}
\end{equation}

\noindent where $\sigma (m)$ is the one sigma uncertainty in the observed magnitude, and $N$ is the number of data points \citep{Montgomery1999}. While the 0.02 magnitude error floor reported by \cite{Jayasinghe2019_III} includes systematic errors, Equation \ref{eq:error} assumes no correlation between individual data points and this assumption may lead to a slight underestimation of the detectable amplitude limit. The photometry for both data sets could likely be improved to an error floor of $\sim$0.005 magnitudes through the use of local differential photometry \citep [e.g.,][]{Mann2011}. Stellar oscillations with amplitudes below the blue line in Figure \ref{fig:precision} are unlikely to be detectable by the current data in each survey. This corresponds to a limit of $\nu_{\rm{max}}\gtrsim1 \; \mu$Hz. 

Figure \ref{fig:HR} shows a modified H-R diagram using surface gravities from \cite{Chaplin2014} and \cite{Yu2018,Yu2020}. The color indicates $\nu_{\rm{max}}$ and the corresponding oscillation period as measured from the \textit{Kepler} data for each star. Figure \ref{fig:HR} shows the well-known relation between the surface gravity of the star and $\nu_{\mathrm{max}}$, which can be expressed as
\begin{equation}
    \frac{g}{g_\odot} = \frac{\nu_{\mathrm{max}}}{\nu_{\mathrm{max},\odot}}\Big(\frac{T_{\mathrm{eff}}}{T_{\mathrm{eff},\odot}}\Big)^{1/2}\:,
    \label{eq:grav}
\end{equation}

 \noindent where $\nu_{\mathrm{max},\odot}$ is 3100 $\mu$Hz, $T_{\mathrm{eff},\odot}$ is 5777 K, and $g_\odot$ is $2.7\times 10^{4} \; \mathrm{cm/s^{2}}$ \citep{Brown1991,Kjeldsen1995}.

Based on our nominal detection threshold in Figure \ref{fig:precision}, the best asteroseismic targets for these ground-based surveys have periods of oscillation longer than $\sim 11$ days and amplitudes greater than $\sim 2.8 \times 10^{3}$ ppm. Figure \ref{fig:HR} shows that these targets are the luminous stars near the TRGB with surface gravities of $\mathrm{log} \ g < 1$.

To test the detectability of oscillations with ground-based transient surveys, we selected 217 red giants from \cite{Stello2014} with measured $\nu_{\mathrm{max}}$ values ranging from 0.20 $\mu$Hz to 5 $\mu$Hz based on \textit{Kepler} long cadence data. This range of asteroseismic observables brackets the range of what should be detectable by ATLAS and ASAS-SN based on Figure \ref{fig:precision} and will allow us to empirically determine the limit to which asteroseismic quantities can be determined.

In addition to the \textit{Kepler} M giants, we use the 16th data release from the APOGEE survey \citep{Ahumada2019}. APOGEE DR16 contains detailed spectroscopic information for a large number of M giants. This spectroscopic data has been used to derive stellar parameters such as metallicity, effective temperature, and surface gravity through spectral synthesis using MARCS model atmospheres. As $\nu_{\rm{max}}$ is directly related to the surface gravity (Equation \ref{eq:grav}), we can compare the asteroseismic surface gravities to those determined spectroscopically. This, along with the comparison to the \textit{Kepler} data, will provide two independent tests of the accuracy with which asteroseismic observables can be determined with ATLAS and ASAS-SN.

\subsection{Ground-Based Surveys}
We utilize photometry from ATLAS and ASAS-SN for our analysis. ATLAS is primarily designed to detect small asteroids on their final approach to Earth. To achieve this ATLAS scans all of the accessible sky every few nights using fully robotic 0.5 m f/2 Wright Schmidt telescopes with a 5.4$\times$5.4 degree field of view. ATLAS began operations with one telescope on Haleakal\=a on the Hawaiian island of Maui in mid 2015, and began operations with their second telescope early 2017 at the Maunaloa Observatory on the big island of Hawai\`{}i. Each ATLAS telescope takes four 30 second exposures per night of $200-250$ target fields covering half of the accessible night sky \citep{ATLAS_original,Tonry2018}. In addition to the search for near-Earth objects, the high-cadence coverage can be used to study variable stars down to a limiting magnitude of $r=18$. The first catalog of variable stars discovered using ATLAS was released by \cite{ATLASvariablestars}. This data release contains observations taken through June of 2017 between a declination of $-30$ degrees and $+60$ degrees. ATLAS uses two customized, wide filters designed to optimize detections of faint objects: the cyan filter ($c$) covering $420 - 650 \; \rm{nm}$ and the orange filter ($o$) covering $560 - 820 \; \rm{nm}$. These filters are well-defined photometric bands described in detail by \cite{ATLAS_original}. While the current data release of variable stars only covers a limited portion of the sky, ATLAS is currently in the process of adding two additional telescopes in the southern hemisphere, providing all-sky coverage in future data releases. 

\begin{figure*}
    \centering
    \includegraphics[width=\textwidth]{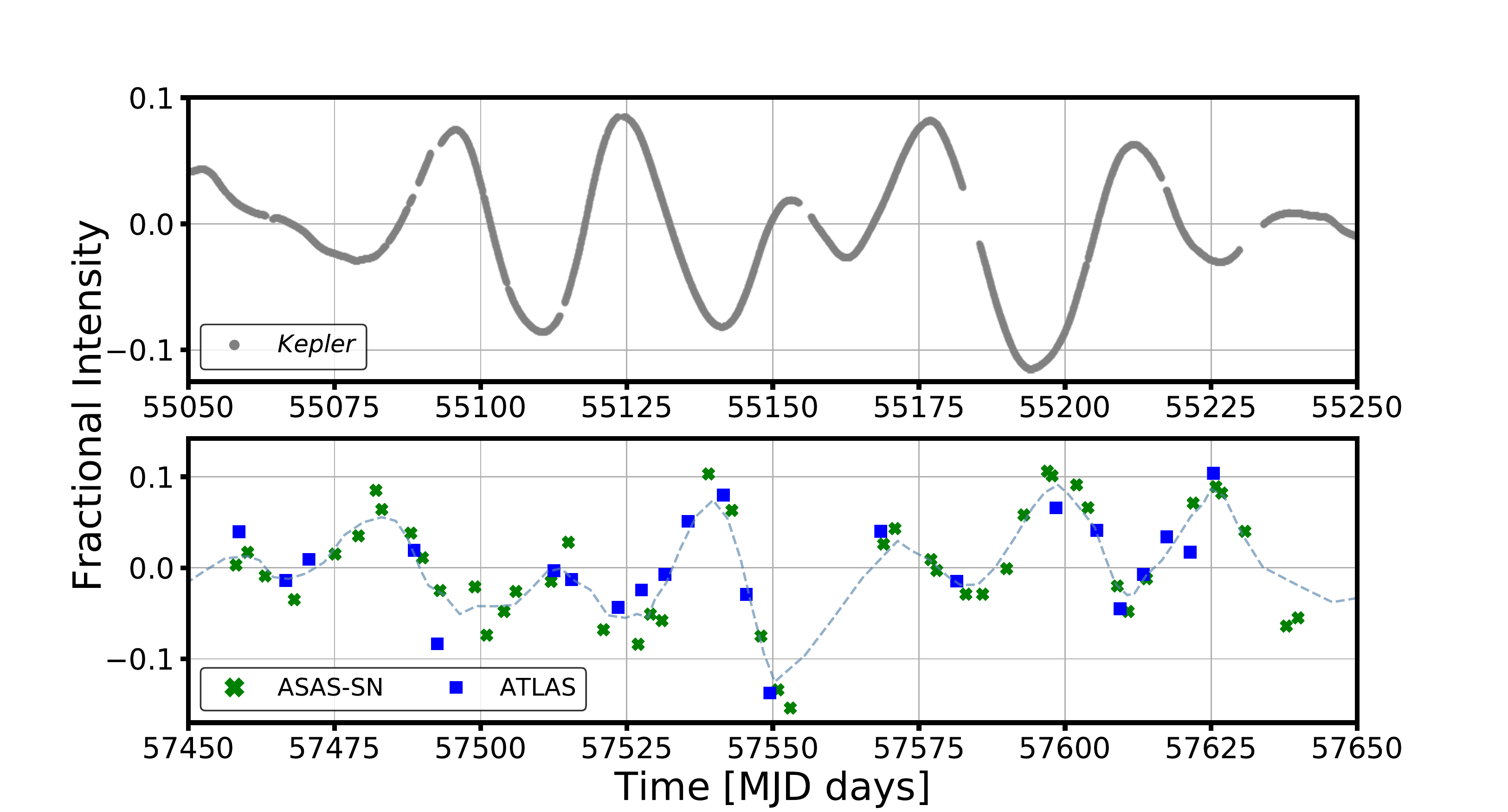}
    \caption{The \textit{Kepler} light curve (top) and the combined ATLAS and ASAS-SN light curve (bottom) (blue squares and green ``x" symbols respectively) of KIC 10976252 (g = 13.8 mag) with a measured frequency of maximum power of 0.4 $\mu$Hz \citep{Stello2014} and a typical period of 29 days \citep{Yu2020}. The flux of the three individual telescopes have been normalized for ease of comparison. The stochastic oscillations of the M giant are clearly visible in both sets of light curves. The dashed line is a rolling average of the ATLAS and ASAS-SN data to highlight the oscillations in the less complete data set.}
    \label{fig:LC}
\end{figure*}

ASAS-SN is the first ground-based survey to monitor the entire visible sky to a depth of $g \sim 18$ mag every night \citep{ASASSN_original,ASASSN_original2}. The primary goal of ASAS-SN is to search for transient objects in order to achieve rapid follow-up. ASAS-SN currently consists of 20 telescopes on 5 mounts located at 4 locations, including one in Hawai\`{}i, two in Chile, one in South Africa, and one in Texas. Each unit consists of four robotic 14 cm telescopes where the field of view of a single ASAS-SN telescope is 4.5$\times$4.5 degrees. Each camera takes three 90 second exposures every epoch, which are then merged to a single image \citep{ASAS-SN_obs}. ASAS-SN began operations in late 2012. The original 2 ASAS-SN units in Hawai\`{}i and Chile covered the entire sky with a cadence of $2-3$ days observing in the Johnson-Cousins $V$-band filter through mid 2018. Three additional units, which were added in late 2017, observe in the SDSS $g$-band filter. These additional units greatly improve the cadence and go one magnitude deeper due to the decreased sky brightness in $g$ compared to $V$. In mid 2018, the original two units were also switched to SDSS $g$-band filters and ASAS-SN currently scans the entire visible sky every $\sim20$ hours to $g \sim 18$ mag. ASAS-SN is also currently adding a 6th unit in China, further increasing the cadence and decreasing sensitivity to weather. Since beginning the search for transients in 2012, ASAS-SN has discovered more than $90,000$ new candidate variable sources and systematically characterized all variables with ASAS-SN light curves \citep{ASASSN_variable,Jayasinghe2019_V,Jayasinghe2019_VI,Jayasinghe2019_II,Jayasinghe2019_III,Pawlak2019_IV}. In this work, we combine these two expansive surveys for the first time to improve sensitivity and show the power that all-sky, ground-based surveys have for the study of stellar variability.
 
The primary filters used by ATLAS and ASAS-SN ($o$ and $g$ respectively) have central wavelengths of $\sim690 \; \rm{nm}$ and $\sim470 \; \rm{nm}$. \cite{Lund2019} showed that the amplitude of oscillation for solar-like oscillators can vary by as much as $15\%$ between different filters based on direct comparisons of oscillations observed by \textit{Kepler} and \textit{TESS}, which have a difference in the central wavelength of their filters of $\sim200 \; \rm{nm}$. A similar amplitude difference is expected between ATLAS and ASAS-SN, though it will not affect the ability to accurately identify asteroseismic observables. The exact strength of the amplitude of oscillations does not affect the values of $\nu_\mathrm{max}$ or $\Delta \nu$ as they are primarily based on the frequency of oscillation. Here, we assume that the amplitude difference will be negligible and simply combine the normalized ATLAS and ASAS-SN data in order in increase the observational baseline and the total number of data points in each light curve.

Figure \ref{fig:LC} shows the light curve of the M giant KIC 10976252 as observed by \textit{Kepler}, ATLAS, and ASAS-SN. This star has an average $g$ magnitude of 13.8 and a measured frequency of maximum power of 0.4 $\mu$Hz, corresponding to a typical oscillation period of 29 days \citep{Stello2014,Yu2020}. The \textit{Kepler} light curve displays clear semi-regular, solar-like oscillations. The combined light curves from ATLAS and ASAS-SN are shown in the bottom panel of Figure \ref{fig:LC}. The small difference in oscillation amplitude can be seen, with the ASAS-SN $g$-band data showing slightly larger oscillations than the ATLAS $o$-band data. While the data from the ground-based surveys is less precise and less continuous than the \textit{Kepler} data, the same semi-regular variations in brightness are clearly recovered. The space and ground-based observations were not taken simultaneously, leading to noticeably different signals. Due to the stochastic nature of solar-like oscillations, the exact frequency and amplitude of the oscillations will vary with time, leading to the different appearance seen in the light curves from the space-based and ground-based data.
\clearpage

\subsection{Asteroseismic Analysis}
We extracted ATLAS and ASAS-SN light curves for 217 red giants from \cite{Stello2014}. We calculated the power spectra for each source utilizing the \texttt{astropy.timeseries} package \texttt{LombScargle} \citep{VanderPlas2012,VanderPlas2015}. Each power spectrum was initially examined by eye to see whether asteroseismic oscillations were present. For a majority of these targets, a single large peak corresponding to the dominant frequency of oscillation was clearly identifiable. For many of the sources no additional oscillation modes could clearly be identified making the determination of the large frequency separation ($\Delta \nu$) impossible. We therefore focused on the ability to accurately determine the frequency of maximum power. 

We measured both $\nu_{\rm{max}}$ and the dominant period of the combined ATLAS and ASAS-SN data. For each of these observables we only analyzed the region of the power spectra that falls below the Nyquist frequency (typically $>$2 $\mu$Hz), defined as $1/(2\Delta t)$ where $\Delta t$ is the separation in time between consecutive data points. As unevenly sampled data do not have a true Nyquist frequency \citep{LombScargle}, we used the average separation of adjacent observations in an estimate of the effective Nyquist frequency. The frequency of maximum power for the stars in our sample that are likely to have detectable oscillations will fall below the Nyquist frequency given the sampling of ATLAS and ASAS-SN. This limit also provides a useful frequency range that can be used to determine the noise of the power spectrum. We note that our frequency analysis is unaffected by aliasing, since the strongest sidelobes from daily gaps in data ($\sim$11 $\mu$Hz) are much larger than the frequency range over which we search for oscillations.

The value of $\nu_{\mathrm{max}}$ and the dominant period for each star was determined using only the ATLAS light curve, only the ASAS-SN light curve, and a combination of the ATLAS and ASAS-SN light curves. The data with the strongest detection was then chosen based on the signal-to-noise (S/N) ratio of the detection and the false alarm probability (FAP) that was reported by the \texttt{LombScargle} package. The noise was determined by the mean power just below the Nyquist frequency and the FAP is the probability that a data set consisting of white noise with no periodic signal could produce a peak of a similar magnitude through coincidental alignment among random errors \citep{LombScargle}. The light curve with the highest S/N and the lowest FAP was used to define the frequency of maximum power and the dominant period. For a majority of the targets that had data in both ground-based surveys, the combined light curves provided the clearest signal (64$\%$), however there were a few cases where either the ATLAS (12$\%$) or ASAS-SN (23$\%$) data alone had a stronger signal than when combined. There were also 60 targets for which only ASAS-SN photometry was available. When determining the precise frequency of maximum power, we smoothed the power spectra using a Gaussian smoothing function with a kernel of $\sigma =  0.05 \; \mu \mathrm{Hz}$ to include the power from any additional modes that might be present. The dominant period of oscillation was simply chosen by taking the highest peak in the power spectrum below the Nyquist frequency. 

\begin{figure*}
    \centering
    \includegraphics[width=\textwidth]{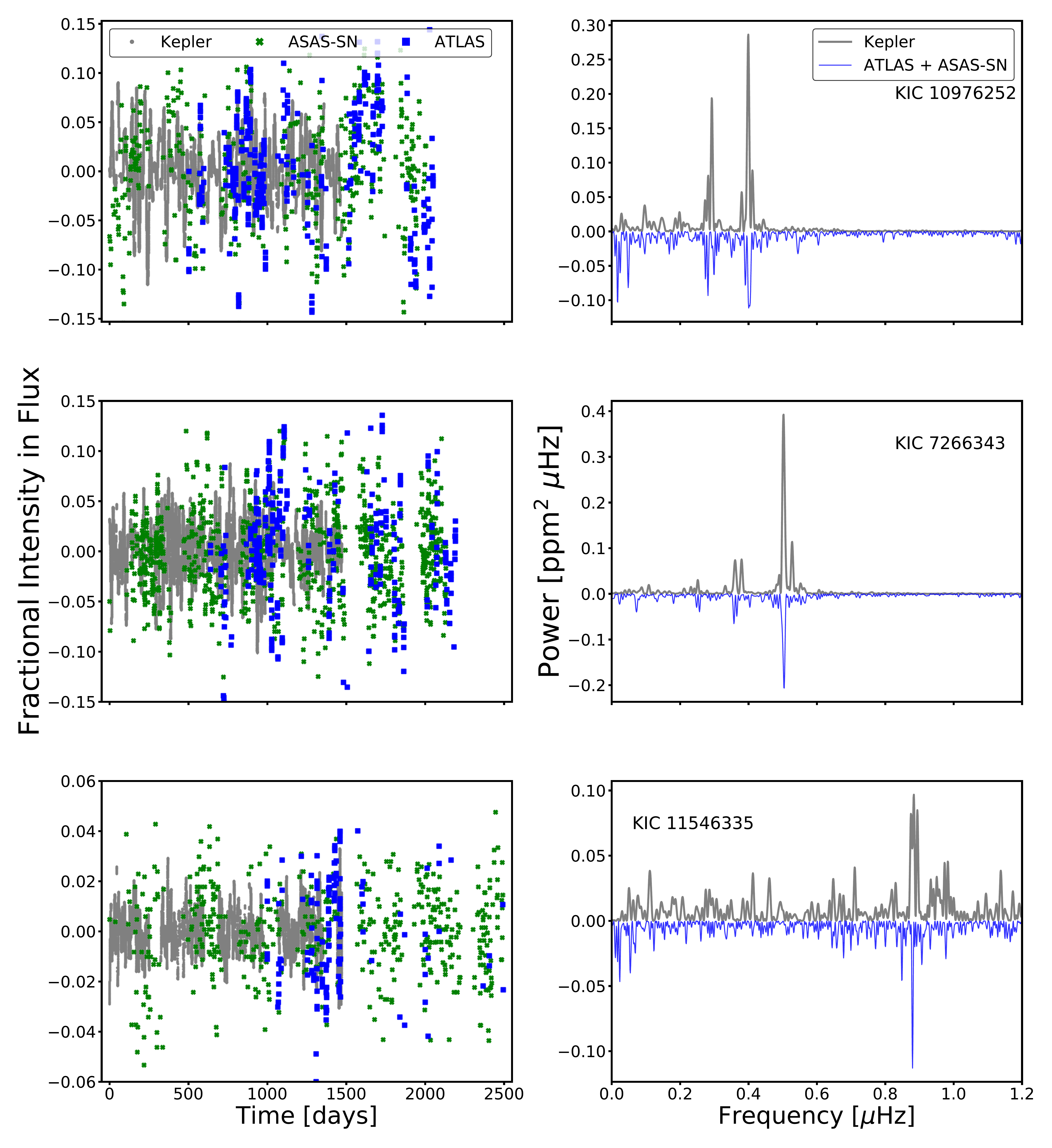}
    \caption{Three examples of \textit{Kepler} light curves (left panels, gray points) and the corresponding power spectra (right panels, gray lines). Over plotted are the ATLAS (blue squares) and ASAS-SN (green ``x" symbols) light curves and the corresponding power spectra from these combined light curves (blue lines). The \textit{Kepler} and ground-based observations for each of these stars were not taken simultaneously, but the light curves have been normalized and over plotted to allow for a qualitative comparison. The power spectra from the ground-based data have been inverted for ease of comparison.}
    \label{fig:LCPS}
\end{figure*}

Figure \ref{fig:LCPS} shows three example light curves where we clearly recovered the same asteroseismic signal from both the \textit{Kepler} data and the combined ATLAS and ASAS-SN data. While the photometric quality of the ground-based data is not as high as the space-based data, the relative baseline of observations for the ground-based data is approximately twice as long, leading to a frequency resolution which is almost twice as high.  

\section{Results}
We analyzed all 217 M giants and then inspected the results to determine the FAP, S/N, peak amplitude, and number of good data points in the light curve that would automatically remove the largest outliers, while minimizing the number of false negatives. The resulting criteria are:

\begin{itemize}
    \item Detections must have a FAP lower than $10^{-10}$.
    \item Detections must have a S/N greater than 25.
    \item Detections must have a peak amplitude in the power spectra greater than 0.1 ppm$^{2}$ $\mu$Hz$^{-1}$.
    \item Detections must have light curves with more than 100 data points falling within 3$\sigma$ of the mean of the light curve and with a reported error lower than 0.05 magnitudes.
\end{itemize}

These criteria work well for M giants with a \textit{Kepler} $\nu_{\mathrm{max}}$ less than 1 $\mu$Hz, as 65.7$\%$ of these stars met these criteria and had a detectable frequency of maximum power with ATLAS and ASAS-SN data. No stars in our sample of 217 M giants with a known frequency of maximum power greater than 1 $\mu$Hz had ATLAS and ASAS-SN data that met the criteria, confirming the estimated detection limit shown in Figure \ref{fig:precision}.

\begin{figure}
    \centering
    \includegraphics[width=\linewidth]{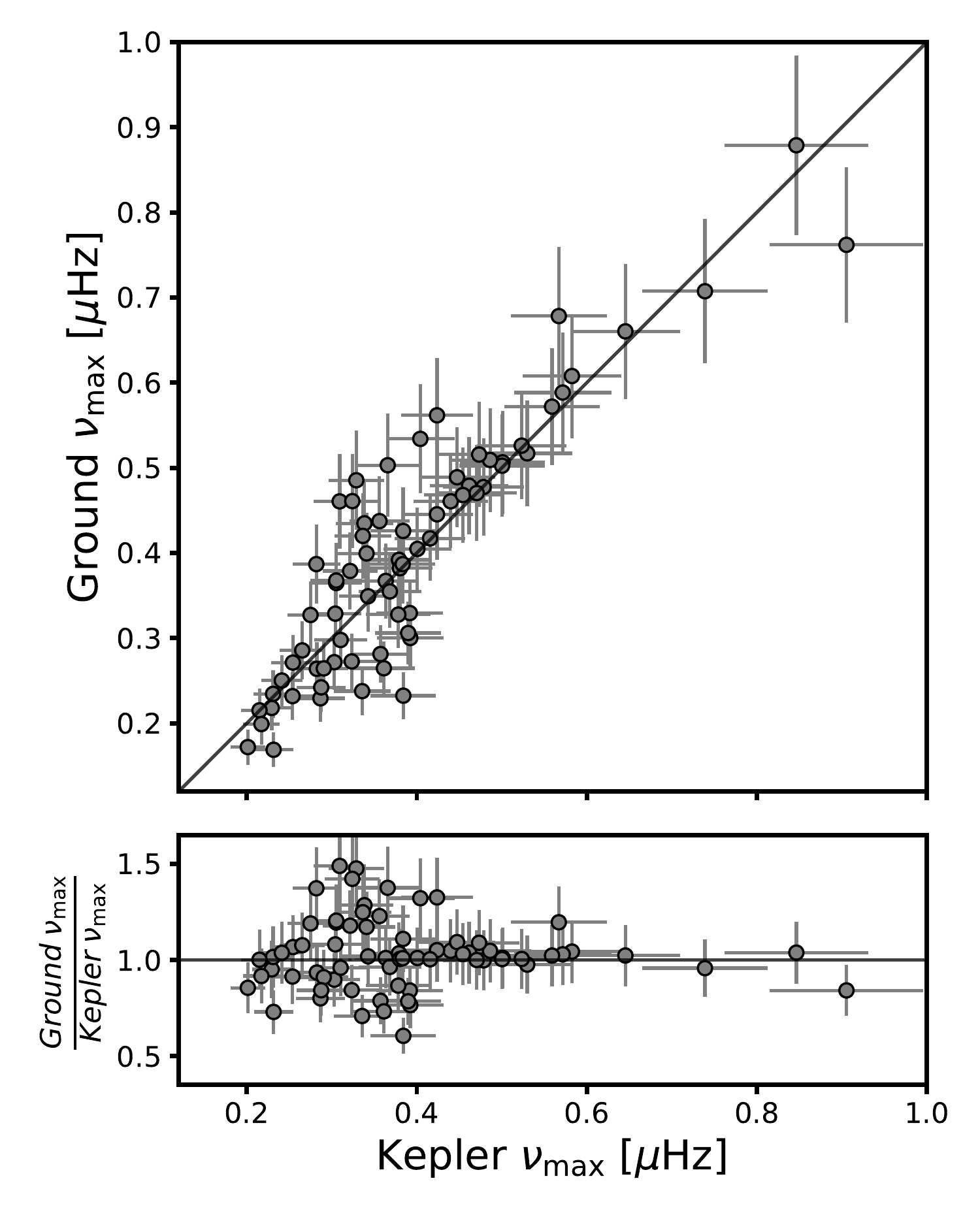}
    \caption{Comparison of $\nu_{\mathrm{max}}$ as measured from \textit{Kepler} \citep{Stello2014} and the ground-based surveys ATLAS and ASAS-SN. The black line shows the one-to-one relation.}
    \label{fig:Scatter}
\end{figure}

Figure \ref{fig:Scatter} shows the comparison of the \textit{Kepler} frequency of maximum power determined by \cite{Stello2014} and that determined with the ground-based data for sources that meet the detection criteria. The errors in the detections from the ground-based data are found through Monte Carlo simulations, where the power spectra for each source is drawn from a chi-square distribution with two degrees of freedom \citep[see e.g.,][]{HuberTestingScaling}. The residuals in the bottom panel of Figure \ref{fig:Scatter} show that the $\nu_{\mathrm{max}}$ determined with ATLAS and ASAS-SN data is in agreement with the $\nu_{\mathrm{max}}$ determined with \textit{Kepler} data with a mean fractional difference of $3 \pm 2\%$ and a scatter of $18\%$. Some of this scatter can be attributed to the fact that solar-like oscillations are a stochastic process that causes $\nu_{\mathrm{max}}$ to moderately vary with time since most of the ATLAS and ASAS-SN data were not taken simultaneously with the \textit{Kepler} data. Table \ref{tab:kepler} lists our asteroseismic results for the \textit{Kepler} sample.

\begin{figure}[t!]
    \centering
    \includegraphics[width=\linewidth]{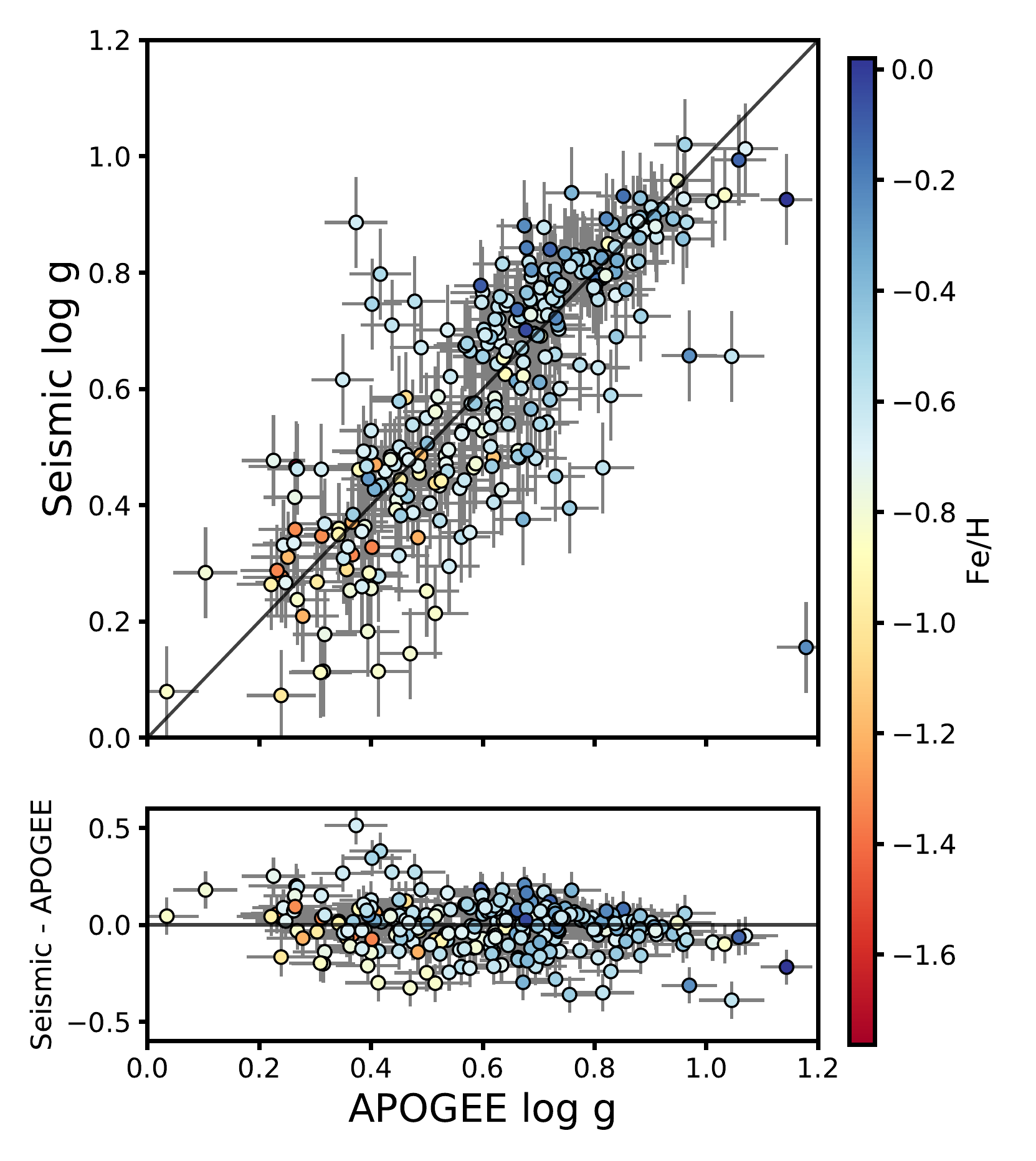}
    \caption{Comparison of the surface gravities for M giants as measured by APOGEE and the asteroseismic analysis of ground-based light curves. The black line represents the one-to-one relation and the color is the metallicity of each source as reported by APOGEE DR16.}
    \label{fig:APOGEE}
\end{figure}

As a second test, we compare asteroseismic surface gravities (calculated using Equation \ref{eq:grav}) to those determined spectroscopically in APOGEE DR16 \citep{Ahumada2019}. To determine the asteroseismic surface gravities we use $\nu_{\rm{max}}$ values found with ATLAS and ASAS-SN data and $T_{\rm{eff}}$ reported by APOGEE DR16. Figure \ref{fig:APOGEE} shows the comparison for 276 randomly selected stars that were present within the ATLAS variable star catalog and had frequencies of maximum power that were detectable with ATLAS and ASAS-SN data. The surface gravity measurements determined from ATLAS and ASAS-SN are in agreement with a mean difference of $0.01\pm0.01 \; \rm{dex}$ and with a scatter of 0.1 dex. This is consistent with the precision of the $\nu_{\mathrm{max}}$ detections using ATLAS and ASAS-SN data for the \textit{Kepler} stars. Table \ref{tab:appogee} lists our asteroseismic results for the APOGEE sample.

\begin{table*}
{\centering
\caption{Asteroseismic Results for the \textit{Kepler} Sample}

\tablenum{1}
\begin{tabularx}{\textwidth}{YYYYYY}
\hline \hline
\textit{Kepler} ID & \textit{Kepler} mag & $\nu_{\rm{max}}$ (ground) & Period (ground) & $\nu_{\rm{max}}$ (\textit{Kepler}) & Period (\textit{Kepler}) \\ 
 &  & [$\mu$Hz] & [days] & [$\mu$Hz] & [days] \\  
\tableline
893210 & 11.94 & 0.437  & 26.050 & 0.356 & 26.54 \\
1430207 & 10.76 & 0.272 & 40.493 & 0.303 & 43.25 \\
1434591 & 11.46 & 0.300 & 41.546 & 0.392 & 29.41 \\
2011145 & 11.52 & 0.506 & 22.895 & 0.501 & 22.90 \\
2141385 & 8.26 & 0.387 & 30.330 & 0.282 & 30.13 \\
2163829 & 10.01 & 0.329 & 28.201 & 0.392 & 28.28 \\
2302624 & 10.46 & 0.232 & 59.496 & 0.384 & 29.53 \\
2421898 & 11.07 & 0.461 & 24.826 & 0.440 & 24.51 \\
2569126 & 11.98 & 0.608 & 18.826 & 0.582 & 17.18 \\
2569935 & 13.12 & 0.328 & 42.357 & 0.378 & 30.26 \\
\tableline
\end{tabularx}}
\tablecomments{The ground-based $\nu_{\rm{max}}$ measurements have a typical fractional error of 12$\%$. This table is available in its entirety (72 rows) in a machine-readable form in the online journal and on the arXiv.}

\tablerefs{Column 5: \cite{Stello2014},
Column 6: \cite{Yu2020}}
\label{tab:kepler}
\end{table*}

\begin{table*}
{\centering
\caption{Asteroseismic Results for the APOGEE Sample}
\tablenum{2}
\begin{tabularx}{\textwidth}{cYYYY}
\hline \hline
APOGEE ID & $\nu_{\rm{max}}$ & log g (seismic) & $T_{\rm{eff}}$ & log g (APOGEE) \\ 
& [$\mu$Hz] & [cgs] & [K] & [cgs] \\
\tableline
2M00313194+4920079 & 0.454 & 0.691 & 3879 & 0.703 \\
2M00355825+5007402 & 0.580 & 0.795 & 3851 & 0.723 \\
2M00363406+5201364 & 0.493 & 0.726 & 3868 & 0.705 \\
2M00405247+5026111 & 0.447 & 0.681 & 3827 & 0.630 \\
2M00445288-1244488 & 0.164 & 0.238 & 3703 & 0.268 \\
2M00494387+3910184 & 0.541 & 0.767 & 3888 & 0.723 \\
2M00501785+3921390 & 0.207 & 0.347 & 3848 & 0.312 \\
2M01022521+5141523 & 0.412 & 0.643 & 3785 & 0.625 \\
2M01095951+5055340 & 0.611 & 0.819 & 3865 & 0.754 \\
2M02245023+4658344 & 0.242 & 0.410∂ & 3748 & 0.446 \\
\tableline
\end{tabularx}}
\tablecomments{Typical uncertainties are 18$\%$ for $\nu_{\rm{max}}$ (see Figure \ref{fig:Scatter}) and $\sim$70 K for $T_{\rm{eff}}$ \citep[see][]{Ahumada2019}. This table is available in its entirety (277 rows) in a machine-readable form in the online journal and on the arXiv.}

\tablerefs{Columns 4 \& 5: \cite{Ahumada2019}}
\label{tab:appogee}
\end{table*}

\begin{figure*}
 \centering
 \includegraphics[width=\textwidth]{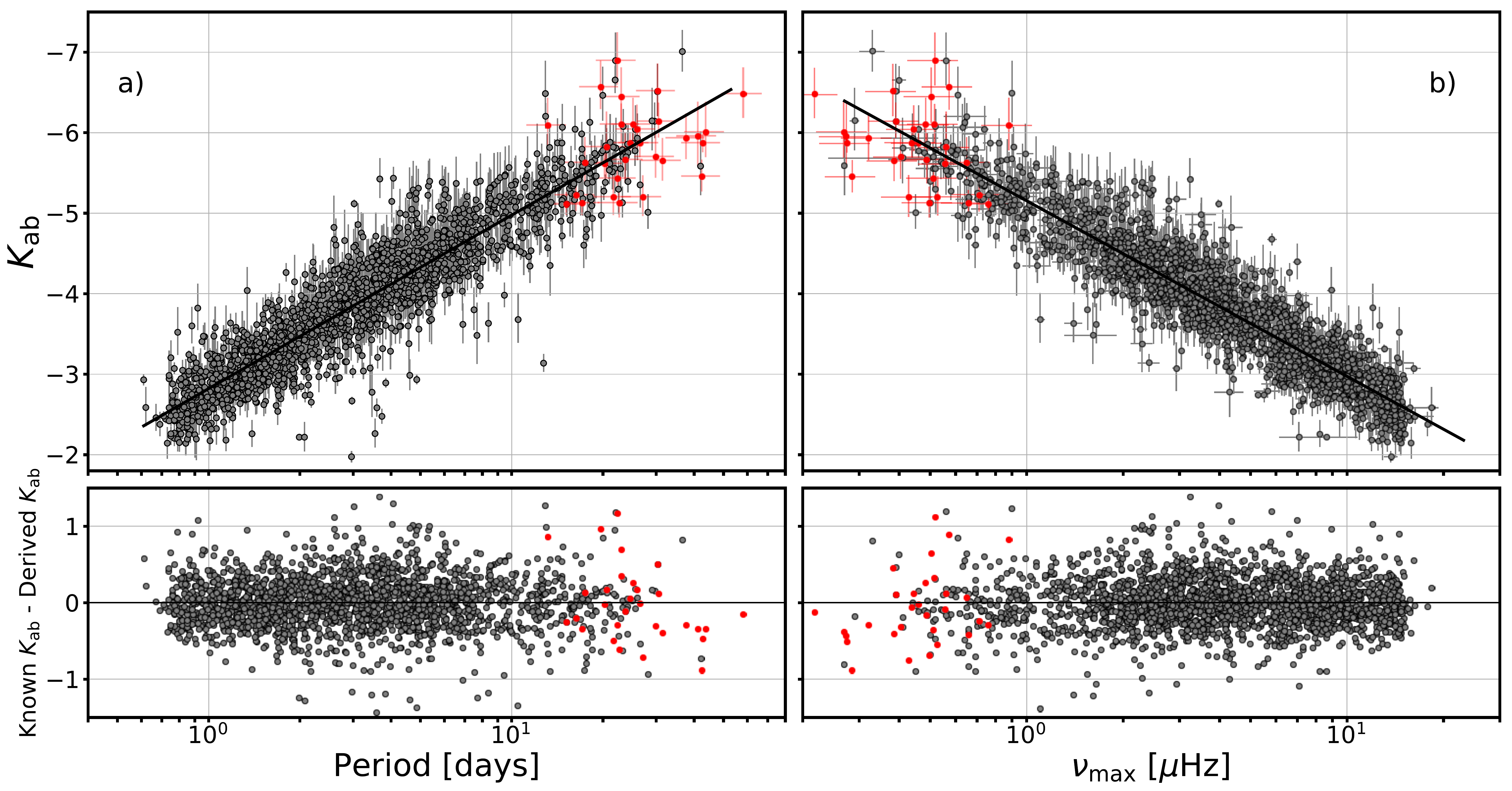}
\caption{Left: Gaia absolute $K$-band magnitudes from \cite{Berger2018} as a function of the dominant period of oscillation reported by \cite{Yu2020} for a sample of red giants in the \textit{Kepler} field (top) and the residual between the known absolute $K$ magnitude reported in \cite{Berger2018} and the absolute $K$ magnitude found from the best-fit line (bottom). Right: Same as the left, but using the frequency of maximum power reported by \cite{Yu2020} instead of the dominant period of oscillation. The red points use the dominant period of oscillation and the frequency of maximum power derived from ATLAS and ASAS-SN data rather than \textit{Kepler} data.}
\label{fig:Kab}
\end{figure*}

The agreement seen in Figure \ref{fig:APOGEE} is remarkable for two reasons. First, the asteroseismic scaling relations are anchored to the Sun and thus not expected to be accurate for evolved stars near the TRGB. Secondly, cooler stars are more poorly fit by model atmospheres due to uncertainties from the increased number of molecular absorption features and non-spherical atmospheres. This may cause spectroscopic surface gravity measurements to be affected by systematic errors, and may be responsible for the slight offset around a surface gravity of $\sim$0.7 dex seen in Figure \ref{fig:APOGEE}. The fact that the surface gravity values agree as well as they do in Figure \ref{fig:APOGEE} suggest that asteroseismic scaling relations for evolved stars are potentially more reliable than previously thought. \cite{Zinn2019}, showed that asteroseismic radii for large stars ($>$30 $R_\odot$) are too large by 8.7$\pm$0.9$\%$ (rand.) and $\pm$2.0$\%$ (syst.) when compared to radii determined using Gaia DR2 data. \cite{Yu2020} finds similar offsets for the radii of large stars and stars with $\nu_\mathrm{max} < 3 \; \mu \mathrm{Hz}$. The strong agreement we find between the APOGEE surface gravities and the asteroseismic surface gravities implies that the differences from past studies may be due to selection effects when using small parallaxes for more distant stars. Alternatively, the scaling relation between $\nu_{\mathrm{max}}$ and the surface gravity used here may be accurate, but the scaling relation used to determine the stellar radius, which requires a combination of $\nu_{\mathrm{max}}$ and $\Delta \nu$, is not. The strong agreement may also imply that the spectroscopic surface gravities reported by APOGEE DR16 are more accurate than the previous data releases used in past studies, thanks to improvements in the model atmospheres in the APOGEE pipeline. 

We note that stars near the TRGB do not span a large temperature range (see Table \ref{tab:appogee}). The narrow temperature range of these stars along with the weak dependence on temperature seen in Equation \ref{eq:grav} allows asteroseismic scaling relations to be used solely with photometric data by assuming an average temperature \citep{Bellinger2020}. Indeed, we have confirmed that using such an approach does not significantly change the agreement in Figure \ref{fig:APOGEE} between our asteroseismic surface gravities and the spectroscopic surface gravities.

\section{Asteroseismic Distances to Galactic M Giants}

\subsection{Absolute Magnitudes from Period-Luminosity Relations}

The existence of a period-luminosity relationship for M giants has been well studied in the past, primarily using light curves obtained for microlensing surveys \citep[e.g.,][]{Kiss2003,Kiss2004,Groenewegen2004,Ita2004,Soszynski2007}. In particular, \cite{Tabur2009} measured the dominant period of oscillation for a sample of red giants from the Large Magellanic Cloud (LMC), the Small Magellanic Cloud (SMC), and the  Galactic Bulge using 5.5 years of ground-based photometric data. \cite{Tabur2010} found multiple ridges in the M giant period-luminosity relation due to different radial orders. By examining the density of stars displaying dominant periods along the different orders, \cite{Tabur2010} derive probability density functions that describe the likelihood of a star having a particular absolute magnitude given the observed frequencies. \cite{Tabur2010} go on to show that the period-luminosity sequence zero-points have a negligible metallicity dependence and emphasize that the RGB pulsation properties are consistent and universal, indicating that the period-luminosity sequences are suitable as distance indicators.

In this work we test the precision of a simpler technique, using only $\nu_{\rm{max}}$ or the single measured dominant period to determine the absolute magnitude of the source. In Figure \ref{fig:Kab} we show the relation between the known absolute $K$-band magnitude for a sample of stars from \cite{Berger2018}, based on Gaia DR2 parallax measurements, and the single dominant period and the frequency of maximum power measured using \textit{Kepler} data \citep{Yu2020} along with the dominant period and $\nu_{\mathrm{max}}$ value determined with ATLAS and ASAS-SN data. A clear relation is seen in each panel: as the absolute $K$-band magnitude becomes brighter, the dominant period of oscillation increases and the $\nu_{\rm{max}}$ decreases. 

\begin{figure*}
    \centering
    \includegraphics[width=\textwidth]{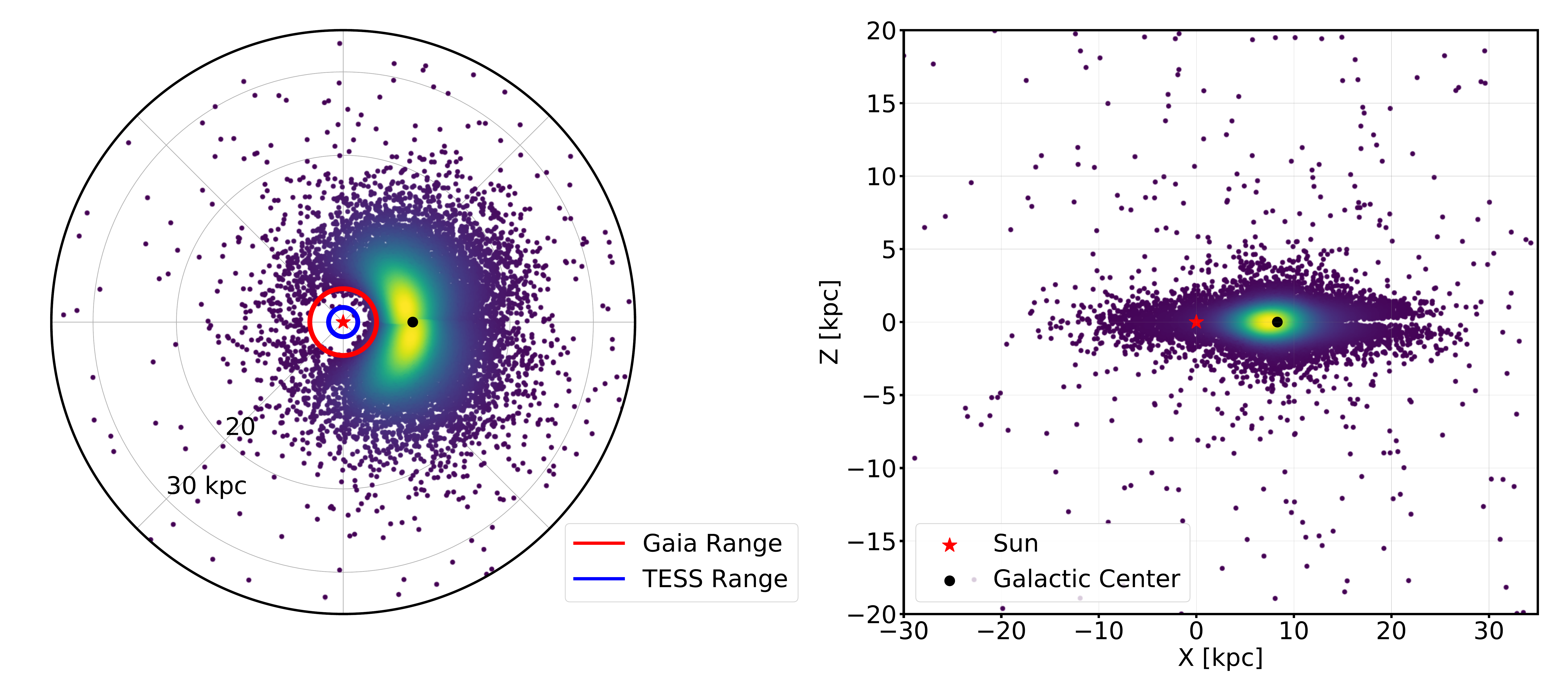}
    \caption{Synthetic population of Milky Way M giants accessible to asteroseismic distance measurements from ground-based transient surveys. Left: A top-down view of the Milky Way, centered at the Sun (red star) with the galactic center shown as a black dot. The number of stars has been down sampled to 1$\%$ of the complete synthetic population. The red circle shows the distance at which Gaia will have a similar precision in distance as that derived from the period-luminosity relation ($\sim15\%$ at $\sim$4 kpc). The blue circle show the distance at which asteroseismic distances can be determined with \textit{TESS} data of red clump stars ($\sim$2 kpc). The lack of stars close to the Sun visible in the left panel is caused by M giants that would be too bright for ATLAS and ASAS-SN to observe being removed. Right: A view through the disk of the Milky Way.}
    \label{fig:SimDis}
\end{figure*}

The bottom panels of Figure \ref{fig:Kab} show the residuals between the absolute $K$ magnitude reported by \cite{Berger2018} and the absolute $K$ magnitude found using the best-fit line to each relation. The standard deviation for each of these residuals is 0.30 magnitudes. Some of the scatter in this relation is due to the precision of the known absolute magnitude measurements, which have a typical error of $\sim$0.16 magnitudes. This uncertainty is dominated by the parallax measurements, which can be uncertain by up to 22$\%$ for these very luminous and distant stars. However, the dominant source of scatter in the period-luminosity relation is likely due to our simplified method, which averages over the individual ridges in the period-luminosity relation, as well as the stochastic nature of the oscillations. 

Given the precision with which the frequency of maximum power can be determined with ATLAS and ASAS-SN data ($\sim$18$\%$), the precision of the absolute $K$ magnitude (and therefore the distance to the star) will be dominated by the scatter in the observed period-luminosity sequence (a maximum of $\sim$30$\%$ in magnitude or $\sim$15$\%$ in distance). \cite{GaiaDR2} quotes the typical uncertainties of Gaia DR2 parallax measurements as 0.04 mas for sources brighter than $G \sim14$ mag, corresponding to an error of $\sim 15 \%$ distance uncertainty at $4 \; \rm{kpc}$. Thus, ATLAS and ASAS-SN can provide more precise distances for the most luminous M giants throughout the Milky Way than can be achieved with Gaia. The precision in distance could potentially be improved further to $\sim10\%$ if a more thorough analysis utilizing the multiple modes of oscillation were conducted \citep{Tabur2010}.

\subsection{Expected Distance Yield}

To predict the approximate yield of an asteroseismic survey utilizing all-sky, ground-based transient surveys we generated a synthetic stellar population for the Milky Way using \texttt{Galaxia} \citep{Sharma2016}. We used the default \texttt{Galaxia} settings, simulating the thin disk, thick disk, and halo components down to V $<$ 19 magnitude including extinction using the standard \texttt{Galaxia} reddening model. The synthetic population was randomly down sampled to 1$\%$ to speed up computation, resulting in a total of $\sim 4 \times 10^{6}$ stars.

From this catalog we selected M giants near the TRGB that would have a $\nu_{\rm{max}} <$ 1 $\mu$Hz, with $\nu_{\rm{max}}$ values calculated from $T_{\rm{eff}}$ and the surface gravity for each star. We also removed stars that have an apparent magnitude that would be too bright ($K \lesssim 8$ mag) or too faint ($K \gtrsim 13$ mag) to be observed by ATLAS and ASAS-SN. This resulted in $\sim2.2 \times 10^{6}$ stars (corrected for the down sampling) which should have oscillations observable with ATLAS and ASAS-SN. We impose no declination restrictions, as ATLAS is in the process of expanding to the southern hemisphere, allowing for all-sky coverage similar to ASAS-SN.

Figure \ref{fig:SimDis} shows the density of the stars that would be potential candidates for distance measurements using asteroseismology with ATLAS and ASAS-SN relative to the Sun. This approach can probe well into the halo, as 250,000 of these stars are located more than 15 kpc from the Sun. By a distance of $\sim$30 kpc, $\sim$50$\%$ of the stars in the simulation are identified as being members of the Galactic halo, and beyond 40 kpc, 100$\%$ of M giant candidates are identified as halo members. The simulation predicts that there should be approximately 70,000 targets for future ground-based asteroseismic studies that are halo stars, allowing for unique studies of the Galactic halo that are not presently possible.

Figure \ref{fig:SimDis} also shows the distance within which a similar asteroseismic analysis could be done using red clump stars with \textit{TESS}. \textit{TESS} is typically limited to red clump stars due to its short observational baseline across most of the sky \citep[e.g.,][]{Aguirre2020}. This restricts the analysis to stars that are much closer to the Sun than the extremely luminous stars on the TRGB that can be probed with the longer temporal baseline of ground-based surveys. Nearby M giants are not accessible with ATLAS and ASAS-SN, as they saturate if located closer than $\sim$5 kpc to the Sun. This distance roughly corresponds to the sphere where Gaia provides more precise distances, allowing for future calibration between these two methods.

Determining precise distances to M giants in the outer regions of the Milky Way has far reaching applications for both Galactic archaeology and dynamics. Current HI maps of the Milky Way use kinematic distances \citep{Levine2006} that may be uncertain. These maps display large perturbations both in the plane and in the vertical direction that may be the gravitational signature of Galactic satellites \citep{Chakrabarti2009}. Accurate distances out to $\sim$20 kpc would allow the HI map of the Milky Way to be re-derived and allow for tests of the interaction models. This would have a significant impact on our understanding of gas dynamics in the Milky Way. In addition, deriving distances out to $\sim$50 kpc in the stellar halo would enable us to obtain constraints on the Galactic potential using action-space clustering \citep{Sanderson2015,Sanderson2016}. Detailed comparisons of dynamical modeling of individual stellar streams of newly discovered dwarf galaxies, such as the Antlia 2 dwarf galaxy \citep{Chakrabarti2019}, which is expected to have had a close approach to the Milky Way, will also yield complementary constraints to the Galactic potential. Both of these methods would substantially improve our understanding of the dynamical evolution of the Milky Way, and hinge crucially on obtaining more accurate distances, which can be provided by asteroseismic distances from ground-based surveys at sufficient precision.

The discovery of stellar streams in the outer halo at distances greater than 100 kpc \citep{Sesar2017} using RR Lyrae can yield important constraints on tidally interacting dwarf galaxies and the Galactic potential. Stars in the distant halo provide constraints on different populations of accreted dwarf galaxies, which can be inferred using the relative numbers of M giants to RR Lyrae stars as a proxy for the accretion time, thereby constraining the accretion history of the Milky Way \citep{Sanderson2017}. Similarly, the relative numbers of M giants and RR Lyrae stars can be used to study the interaction history of the outer Galactic disk \citep{Price-Whelan2015}. The complementary (and in some cases crucial) information that can be obtained from the combination of classical pulsators and M giants is another example of the applicability of reliable asteroseismic distances towards better understanding the history and dynamics of the Milky Way. 

\section{Conclusions}

We have used light curves of M giants from the ground-based transient surveys, ATLAS and ASAS-SN to study their potential as asteroseismic distance indicators. Our main conclusions are as follows:

\begin{enumerate}
    \item ATLAS and ASAS-SN can recover oscillations of M giants with a frequency of maximum power ($\nu_{\rm{max}}$) less than $\sim$1 $\mu$Hz. This corresponds to M giants with luminosities of $\gtrsim$650 L$_\odot$ and oscillation periods of $\gtrsim$11.6 days.
    
    \item The recovered $\nu_{\rm{max}}$ values agree with the \textit{Kepler} values to 18$\%$ and the derived surface gravity values agree with APOGEE surface gravities to $\sim$0.1 dex. This implies that asteroseismic and/or spectroscopic surface gravity measurements for M giants are less biased than previously thought. If our asteroseismic surface gravity measurements are accurate, this provides an independent validation of spectroscopic surface gravities for M giants. 
    
    \item M giant distances using a simple period-luminosity relation are precise to approximately 15$\%$. This precision could be improved to $\sim10\%$ by utilizing a more in-depth period-luminosity relation using multiple oscillation frequencies \citep[e.g.,][]{Tabur2010}. 
    
    \item Based on all-sky synthetic stellar populations, we estimate that the current ground-based transient surveys hold the potential to measure distances to $\sim2.2 \times 10^{6}$ M giants with a percision of $\sim10-15\%$, 250,000 of which are located more than $15 \; \rm{kpc}$ from the Sun and 70,000 that belong to the galactic halo. This vastly improves over Gaia, which can only achieve distances with a 15$\%$ precision out to approximately 4 kpc.
\end{enumerate}

Our results demonstrate the powerful potential of using ground-based asteroseismic studies of M giants for Galactic archaeology and galactic dynamic studies. Ground-based surveys such as ATLAS and ASAS-SN will continue to collect data for years to come, providing longer baseline observations which will allow for asteroseismic observables to be determined more accurately for a larger number of stars. There is also significant potential to improve the photometric precision of each of these surveys by utilizing local differential photometry which would expand the number of stars for which accurate asteroseismic observables could be determined. Additionally, detrmining the proper scale factor between the different observational filters could potentially lower noise and result in more detections. Utilizing data from other ongoing ground-based surveys, such as Pan-STARRS and ZTF, could also potentially improve the analysis by contributing more data points to each light curve and therefore reducing noise within the power spectra. The addition of future ground-based surveys, such as the Vera Rubin Observatory \citep[LSST,][]{Ivezi2019}, will also contribute to the diversity of stars such studies can be applied to. Combined, these surveys will provided unprecedented constraints on the distances of stars throughout our galaxy.

\acknowledgments
We thank the referee for the helpful report. We thank Jie Yu and Douglas Compton for providing $\nu_{\rm{max}}$ measurements for \textit{Kepler} stars, and Michael Tucker, Gagandeep Anand, and Anna Payne for useful conversations about the project. 

D.H. acknowledges support from the Alfred P. Sloan Foundation. D.H., S.C., and R.E.S. acknowledge support from the Research Corporation for Science Advancement through Scialog award $\#$26996. 

This work has made use of data from the Asteroid Terrestrial-impact Last Alert System (ATLAS) project. ATLAS is primarily funded to search for near earth asteroids through NASA grants NN12AR55G, 80NSSC18K0284, and 80NSSC18K1575; byproducts of the NEO search include images and catalogs from the survey area.  The ATLAS science products have been made possible through the contributions of the University of Hawaii Institute for Astronomy, the Queen's University Belfast, the Space Telescope Science Institute, and the South African Astronomical Observatory.

We thank the Las Cumbres Observatory and its staff for its continuing support of the ASAS-SN project. ASAS-SN is supported by the Gordon and Betty Moore Foundation through grant GBMF5490 to the Ohio State University and NSF grant AST-1515927. Development of ASAS-SN has been supported by NSF grant AST-0908816, the Mt. Cuba Astronomical Foundation, the Center for Cosmology and AstroParticle Physics at the Ohio State University, the Chinese Academy of Sciences South America Center for Astronomy (CASSACA), the Villum Foundation, and George Skestos.

BJS, KZS, and CSK are supported by NSF grants AST-1515927, AST-1814440, and AST-1908570. BJS is also supported by NSF grants AST-1920392 and AST-1911074.  Support for G.P. is provided by the Ministry of Economy, Development, and Tourism’s Millennium Science Initiative through grant IC120009, awarded to The Millennium Institute of Astrophysics, MAS. TAT is supported in part by NASA grant 80NSSC20K0531. S.C. greatefully acknowledges support from NASA ATP NNX17AK90G, and from Research Corporation for Scientific Advancement’s Time Domain Astrophysics Scialog. DS acknowledges support from the Australian Research Council.

\bibliography{References}
\end{document}